\documentclass[aps,prb,twocolumn,floats,showpacs,superscriptaddress]{revtex4-1}
\usepackage{graphicx,epsfig}
\usepackage{times,bbm}
\usepackage{graphics,dcolumn,bm,float}
\usepackage{amssymb,amsmath,rotate,color}
\usepackage[title,titletoc,toc]{appendix}
\usepackage{mathtools}
\usepackage{booktabs}
\usepackage{tcolorbox}
\usepackage[mathscr]{eucal}
\usepackage{graphicx,epsfig}
\usepackage[normalem]{ulem}
\usepackage{caption}
\usepackage{graphicx}
\usepackage[pagebackref=false,colorlinks,linkcolor=blue,citecolor=blue,urlcolor=magenta]{hyperref}

\begin{document}
\unitlength 1 cm
\newcommand{\be}{\begin{equation}}
\newcommand{\ee}{\end{equation}}
\newcommand{\bea}{\begin{eqnarray}}
\newcommand{\eea}{\end{eqnarray}}
\newcommand{\nn}{\nonumber}
\newcommand{\vk}{\vec k}
\newcommand{\vp}{\vec p}
\newcommand{\vq}{\vec q}
\newcommand{\vkp}{\vec {k'}}
\newcommand{\vpp}{\vec {p'}}
\newcommand{\vqp}{\vec {q'}}
\newcommand{\bk}{{\vec k}}
\newcommand{\bp}{{\bf p}}
\newcommand{\bq}{{\bf q}}
\newcommand{\br}{{\bf r}}
\newcommand{\bR}{{\bf R}}
\newcommand{\up}{\uparrow}
\newcommand{\down}{\downarrow}
\newcommand{\cdag}{c^{\dagger}}
\newcommand{\hlt}[1]{\textcolor{red}{#1}}
\newcommand{\ba}{\begin{align}}
\newcommand{\ea}{\end{align}}
\newcommand{\la}{\langle}
\newcommand{\ra}{\rangle}
 \title{Induced superconductivity in Fermi arcs}
 
    \author{Z. Faraei}
   \email{zahra.faraei@gmail.com}
   \affiliation{Department of Physics, Sharif University of Technology, Tehran 11155-9161, Iran}

 \author{S. A. Jafari}
 \email{akbar.jafari@gmail.com}
 \affiliation{Department of Physics, Sharif University of Technology, Tehran 11155-9161, Iran}
 \affiliation{Center of excellence for Complex Systems and Condensed Matter (CSCM), Sharif University of Technology, Tehran 1458889694, Iran}
\begin{abstract}
When the interface of a superconductor (SC) with Weyl semimetal (WSM) supports Fermi arcs, the chirality blockade eliminates the
induction of superconductivity into the bulk. This leaves the Fermi arc states as the only low-energy degrees of freedom in the proximity problem.
Therefore the SC$|$WSM system will be a platform to probe transport properties that {\it only} involve the Fermi arcs. With a boundary condition 
that flips the spin at the boundary, we find a $Z_2$ protected Bogoliubov Fermi contour (BFC) around which the Bogoliubov quasi-particles disperse
linearly. The resulting BFC and excitations around it leave a distinct $T^2$ temperature dependence in their contribution to specific heat. 
Furthermore, the topologically protected BFC being a Majorana Fermi surface gives rise to a zero-bias peak the strength of which characteristically depends on the 
length of Fermi arc and tunneling strength. 
For the other BC that flips the chirality at the interface, instead of BFCs we have Bogoliubov-Weyl nodes whose location depends 
on the tunneling strength.

\end{abstract}

\maketitle

\section {Introduction} 
\label{int}
One of the interesting implications of gapless three dimensional Weyl semimetals~\cite{Murakami,BurkovHookBalents,Savrasov,exp_Weyl,exp_TaAs} (WSM), with topologically protected band touching points~\cite{BurkonBalents,Bernevig,HalaszBalents,KaneMeleRappe,Carpentier}, is the realization of Fermi arc that is yet protected~\cite{Savrasov,Ojanen,Vishwanath_arc,TaAs_arc2}. ARPES is an appropriate method to observe the Fermi arc shapes~\cite{TaAs_arc2,Hassan_arc,TaAs_arc}, but it is desirable to find the clues of these Fermi arcs in transport experiments.  The problem is that when it comes to transport properties, the Weyl cones in the bulk and Fermi arcs states both being gapless, will jointly contribute to the transport and therefore separation of the bulk degrees of freedom is challenging. One of the methods for separating the arc contribution is to address the superconducting proximity effect. In proximity effect, only the end layers of materials are involved. Therefore bringing a superconductor (SC) to form SC$|$WSM junction, 
even at this simple-minded level of argument, one expects to observe the Fermi arc dominated effects. As we will argue in this paper, yet there is a more fundamental reason 
that makes SC$|$WSM a genuine Fermi arc dominated system for transport purposes. 

In recent years, there have been many studies in the issue of superconducting proximity in WSMs~\cite{}, most of which have been focused on the proximity effect on the bulk states. It has been shown that, as long as the conical dispersion -- related the bulk excitations of WSMs -- is considered, in the vicinity of conventional s-wave superconductors, the Andreev reflection will be blocked: 
Bovenzi \textit{et al}~\cite{chirality blockade}, have shown that in the proximity with conventional s-wave superconductors, if the (momentum) vector connecting 
the Weyl nodes has a component parallel to the interface, the Andreev reflection in a magnetic WSM will be suppressed by the phenomenon of \textit{chirality blockade}. 
The simple explanation of chirality blockade rests on the bulk Hamiltonian of the $\chi\vec\sigma.\vec p$ form. In proximity to spin-singlet
superconductor, the Andreev reflected hole is required to reverse both spin and physical momentum. But physical momentum reversal is 
accompanied by chirality flip upon which the energy $\chi\vec\sigma.\vec p$ can not be conserved. This will block Andreev reflections involving both 
momentum {\it and} spin reversal~\cite{chirality blockade}. However, if the superconductor is not spin-singlet~\cite{Aji}, or the Cooper pairs do not
have zero center of mass momentum (i.e. FFLO superconductivity)~\cite{Kim}, the chirality blockade can be relaxed. Other
possible ways to escape the chirality blockade would involve pseudo-scalar superconductivity~\cite{Faraei2017,Salehi2016,Specular,Rao2,Madsen}.
Since the presence of boundary breaks the inversion symmetry, another 
important situation that relaxes the chirality blockade is the presence of a boundary itself~\cite{asano}.

When the vector connecting the two Weyl nodes is perpendicular to the interface, there would be no chirality
blockade~\cite{chirality blockade}. In this situation there would be no Fermi arcs as well. 
But once the vector connecting the Weyl nodes develops a small component along the interface (the
projection of which is precisely the Fermi arc), the chirality blockade will start to appear. 
The conclusion will be that,
the chiraltiy blockade of bulk degrees of freedom in proximity with conventional superconductors, 
crucially depends on the existence of Fermi arcs. 
So even if the superconducting coherence length is long enough to reach the deep interior in non-Weyl systems,
by chirality blockade in WSMs, the Fermi arc wins the competition and the bulk states will have no contributions in induced superconductivity. 
In this way, the response of a WSM to the proximity with a conventional s-wave superconductor selectively couples to the Fermi arcs only. 
From this point of view, the proximity with conventional superconductors can be regarded as a tool to study the transport properties where
the only relevant low-energy degrees of freedom are the Fermi arc states. So our proposal in this work is to promote SC$|$WSM heterostructure into a platform
to study the Fermi arcs only. Motivated by this, we study the induced superconductivity in SC$|$WSM system, and find more interesting results 
than we expected, namely a topologically protected Bogoliubov Fermi contour (BFC) or Bogoliubov-Weyl (BW) nodes. 

Let us start by reviewing the existing literature on the combination of superconductivity with WSM.  
First class of works start by a Weyl system which is superconductor in the {\it bulk} and examine
the resulting surface states~\cite{Aji,Zhou,Huang}. In this class of works, doping a WSM converts 
the flat band along the nodal direction to crossing flat bands~\cite{Lu_flatbands}. This can be understood
in terms of the non-trivial monopole charge of the the Cooper pairs~\cite{Haldane_nodal FC}.
In this work we are not concerned with this class. 
The second class however, deals with the induction of superconductivity in WSMs and their surface states (Fermi arc states). 
Numerical result of Khanna and coworkers indicates that the Fermi arc states are gapped except for 
the two points corresponding to the projection of Weyl nodes on the surface supporting the Fermi arcs~\cite{Rao2}.

The presence of gapless points in the excitations of Bogoliubov quasi-particle spectrum of WSM can be understood as follows:
The gap term has to be proportional to $\psi^*_R\psi_L$~\cite{Schwartz} which means a term like $\psi^*_R\psi^*_R$.
This is because the complex conjugation exchanges the chirality~\cite{Schwartz}. But this type of terms are
forbidden if one requires zero center of mass Cooper pairs~\cite{RaoReview}. Therefore the spectrum of Bogoliubov excitations
in WSM can not be entirely gapped, and there should exists nodal points or nodal lines. 

In this work we will develop analytical understanding of the induced superconductivity
in Fermi arc states of undoepd WSMs. The analytical approach of present work based on 
classification of boundary conditions (BCs) in WSMs~\cite{Faraei2018,Falko,Beenakker_Boundary cond} and their Greens' function~\cite{Faraei2018}
will enable us, depending on the BC to obtain elliptic BFC or BW nodes. We find that for first type BC that flips the spin at
the boundary, the BFC is protected by a $Z_2$ index and find an appropriate Pfaffian that changes sign across the BFC. 
For this type of BC, the Bogoliubov excitations around BFC 
are linearly dispersing, and therefore contribute a specific heat term that can be distinguished from bulk contributions.
For second type BC that flips the chirality at the boundary, instead of robust BFC, we find
pairs of BW nodes that disperse by changing the tunneling strength. 

The paper is organized as follows:  In section II we adjust our previously developed Green's 
function method for problems involving the superconductivity. In section III we bring the
SC into proximity with WSM and corresponding to two classes of BCs we obtain the nature of
superconductivity induced into Fermi arc states. In section IV we talk about pairing symmetry
and discuss the Majorana character of BFC. We end the paper by summary of main findings in section V.
Details of algebra are presented in the appendix. 

\section{Green's function method}
\label{secII}
\subsection{Green's function for electrons}
In our previous work~\cite{Faraei2017} we have calculated the Green's function
of a normal WSM. Since the present work will be based on our earlier work, let us
briefly summarize its core results. 
For a semi-infinite inversion symmetric WSM with two nodes at $\pm \vec b$ and a hard wall boundary~\cite{Falko} at $z=0$,
the wave equation is,
\bea
\big[ i \hat{\tau}_z \otimes (\vec \sigma \cdot \vec \nabla) + \hat{\tau}_0 \otimes (\vec \sigma \cdot \vec b) + \check{M} \delta(z) \big] \Psi 
= E \Psi,
\label{H.eqn}
\eea
where $\check M$ is a $4 \times 4$ Hermitian, unitary matrix and effectively incorporates the confinement potential at the boundary. 
Pauli matrices $\tau$ and $\sigma$ operate in chirality and spin spaces. 
We work in units of $\hbar=1$. Furthermore, the lengths and velocities are measured in units of $|\vec b|^{-1}$ and $v_F$, respectively. 
Consistency with the constraint of hard wall assumption, gives the following form for the boundary matrix $\check M$s~\cite{Faraei2017}:
\bea
\check{M}= (\cos \gamma) \check{M}_1 +(\sin \gamma) \check{M}_2,
\label{M.eqn}
\eea
where 
\bea
\check{M}_1=
\begin{pmatrix}
0 & e^{- i \Lambda} & 0 & 0\\
e^{ i \Lambda} & 0 & 0 & 0 \\
0 & 0 & 0 & e^{- i \xi}\\
0 & 0 & e^{ i \xi} & 0
\end{pmatrix},
\label{M_1.eqn}
\eea
rotates the in-plane component of the spin through angels $\Lambda=-\cot^{-1} (b_y/b_x)$ and $\xi=\Lambda - \pi$ for the left and right handed electrons, respectively and 
\bea
\check{M}_2=
\begin{pmatrix}
0 & 0 & e^{- i \alpha} & 0\\
0 & 0 & 0 & e^{- i \beta} \\
e^{ i \alpha} & 0& 0 & 0 \\
0 & e^{ i \beta} & 0 & 0
\end{pmatrix},
\label{M_2.eqn}
\eea
which is diagonal in spin space, but mixes the chirality components. 
Independent of the value of $\gamma$, the BC Eq.~\eqref{M.eqn} frightfully reproduce a Fermi arc on 
the surface state that connects the projections of Weyl nodes on the surface~\cite{Faraei2018}.
Requiring the Fermi arc (ray) emitted from one node to end precisely at the other node gives, $\alpha-\beta=\Lambda-\xi$.

For the Hamiltonian Eq.~\eqref{H.eqn}, the electronic single particle Green's function is obtained as:
\bea
\label{G.eqn}
G_{\chi \chi'}^{ \bar\sigma \sigma}(z,z')&=&C_{\chi \chi'}^{ \bar\sigma \sigma}(z') e^{-(q_\chi+i\chi b_z ) z}\\
\nn
&-&\frac{ \chi(k_x^\chi + i \sigma k_y^\chi) }{8\pi^2(q_\chi+i\chi b_z)}e^{-(q_\chi+i\chi b_z) |z-z'|} \delta_{\chi \chi'},
\eea
and
\bea
G_{\chi \chi'}^{\sigma \sigma}(z,z')&=&  \frac{\varepsilon- i \chi \sigma \partial_z + \sigma b_z}{\chi ( k_x^\chi + i\sigma k_y^\chi) } G_{\chi \chi'}^{\bar{\sigma} \sigma }(z,z').
\eea
where $\chi,\chi'=\pm 1$ is the chirality, $\sigma=\pm 1$ represents the spin direction, $\bar{\sigma}=-\sigma$ and
$q_\chi=(k_x-\chi b_x)^2 + (k_y - \chi b_y)^2 - \epsilon^2$. $\epsilon$ is the electron's energy.

The coefficients $C_{\chi \chi'}^{\sigma \sigma'}$ depend on the BC. For $\check{M}_1$-type BC ($\gamma=0$) we have,
\bea
C_{\chi \chi'}^{\bar{\sigma} \sigma}&=&
\frac{\varepsilon- i \chi \sigma q_\chi + 2 \sigma b_z - \chi e^{- i \sigma \theta_\chi} (k_x^\chi + i \sigma k_y^\chi)}{\varepsilon+ i \chi\sigma q_\chi - \chi e^{- i \sigma \theta_\chi} (k_x^\chi + i\sigma k_y^\chi)}\nn\\
&\times& \frac{\chi (k_x^\chi + i\sigma k_y^\chi)}{8\pi^2(q_\chi + i \chi b_z)}
e^{-(q_\chi+i \chi b_z)z' } \delta_{\chi \chi'}
\eea
while for $\check{M}_2$-type BC ($\gamma=\pi/2$) one obtains,
\bea
C_{\chi \chi}^{\bar{\sigma} \sigma}
=\frac{i \chi \sigma  (k_x^\chi+ i\sigma k_y^\chi) (k_x^{\bar{\chi}}+ i\sigma k_y^{\bar{\chi}})}{8\pi^2D_{\chi \chi}^{\bar{\sigma} \sigma}} 
 e^{-(q_\chi+i \chi b_z)z'}, 
\eea
\bea
C_{\bar{\chi} \chi}^{\bar{\sigma} \sigma}
=\frac{\chi  (k_x^\chi+ i\sigma k_y^\chi)}{8\pi^2(q_\chi + i \chi b_z)} 
\big( \frac{N_{\chi \chi}^{\bar{\sigma} \sigma}}{D_{\chi \chi}^{\bar{\sigma} \sigma}} \big) 
 e^{-(q_\chi+i \chi b_z)z'}, 
\eea
where
\bea
D_{\chi \chi}^{\bar{\sigma} \sigma}&=&\bar{\chi} e^{i\chi \theta_{\bar{\sigma}}}(\varepsilon+ i \chi \sigma q_\chi ) (k_x^{\bar{\chi}} + i \sigma k_y^{\bar{\chi}})\nn\\
&-& \chi e^{i\chi \theta_{\sigma}}(\varepsilon+ i \bar{\chi} \sigma q_{\bar{\chi}} ) (k_x^{\chi} + i \sigma k_y^{\chi}),
\eea
and
\bea
N_{\chi \chi}^{\bar{\sigma} \sigma}= D_{\chi \chi}^{\bar{\sigma} \sigma}+ 2 i  \sigma e^{i\chi \theta_{\bar{\sigma}}}  (q_\chi + i \chi b_z)   (k_x^{\bar{\chi}} + i \sigma k_y^{\bar{\chi}}).
\eea
Both BCs produce a Fermi ray (meaning that the shape of Fermi arc is a straight line) connecting 
the projection of Weyl nodes on the surface whose slope is solely determined by vector $\vec b$ as $\tan^{-1}(b_y/b_x)$.

\subsection{Green's functions for holes}
To incorporate superconductivity into our Green's function formulation, we need to 
augment the Green's functions into the Nambu space. So we need the Green's function for the holes as well. 
The electron and hole Hamiltonians are related by the operation of time reversal operator~\cite{chirality blockade}:
\be
H_h(\vec k)= \sigma_y H_e^*(-\vec k) \sigma_y.
\ee
For the Weyl Hamiltonian in Eq.~\eqref{H.eqn}, the corresponding hole Hamiltonian becomes,
\bea
H_h(\vec k)= \tau_z (\vec{\sigma}.\vec k) - \tau_0 (\vec{\sigma}.\vec b), 
\eea
which can be combined with the electronic part to give the Bogoliubov-De Gennes Hamiltonian,
\bea
H_W=
\begin{pmatrix}
H_e & 0\\
0 & - H_h
\end{pmatrix}
\eea

The crucial point in constructing the Green's function for holes is that the particle-hole transformation 
should also operate on matrix $\check M$ in Eq.~\eqref{H.eqn} that encodes the BC information. 
Starting with BC matrix $\check M_1$ of electrons, 
\bea
\check{M}_1 &=& \frac{\hat{\tau}_0+\hat{\tau}_z}{2}\otimes (\cos\Lambda~\hat{\sigma}_x + \sin\Lambda ~\hat{\sigma}_y )\nn\\
&+& \frac{\hat{\tau}_0-\hat{\tau}_z}{2}\otimes (\cos\xi~ \hat{\sigma}_x + \sin\xi~ \hat{\sigma}_y)\nn
\eea
for holes we obtain, $\sigma_y \check{M}_1^* \sigma_y=-\check{M}_1$ which 
is eventually equivalent to the substitution $\Lambda\rightarrow \pi+\Lambda$ and $\xi\rightarrow \pi+\xi$.
This is quite intuitive, as the reflection of an electron with its in-plane spin rotated by angle $\Lambda$ 
after the TR operation can be equivalently viewed as rotation of the spin of a hole by angle $\pi+\Lambda$.
Similarly for $\check M_2$-type BC we have,
\bea
\check{M}_2 &=& (\cos\alpha~\hat{\tau}_x + \sin\alpha ~\hat{\tau}_y ) \otimes \frac{\hat{\sigma}_0+ \hat{\sigma}_z}{2}\nn \\
&+& (\cos\beta~\hat{\tau}_x + \sin\beta ~\hat{\tau}_y ) \otimes \frac{\hat{\sigma}_0-\hat{\sigma}_z}{2} \nn
\eea
which upon particle-hole transformation becomes,
\bea
\sigma_y \check{M}_2^* \sigma_y &=& (\cos\alpha~\hat{\tau}_x - \sin\alpha ~\hat{\tau}_y ) \otimes \frac{\hat{\sigma}_0- \hat{\sigma}_z}{2}\nn \\
\nn
&+& (\cos\beta~\hat{\tau}_x - \sin\beta ~\hat{\tau}_y ) \otimes \frac{\hat{\sigma}_0+\hat{\sigma}_z}{2} \nn
\eea
Therefore the $\check M_2$ BC matrix for holes is obtained from the corresponding $\check M_2$ of
electrons by the replacement $\alpha\leftrightarrow -\beta$. 

Now, we are ready to set up the Green's function for holes. For this we need to solve 
$$[\varepsilon+H_h+\check M_h\delta(z)] G_h=\delta(\vec r- \vec r')$$ where the matrix $\check M_h$ can 
be any of the matrices discussed above. 
Another important technical point is that the hole part of the wave function is,
$$\psi_h=\big[ -\psi^*_{+\downarrow}~,~ \psi^*_{+\uparrow}~,~-\psi^*_{-\downarrow}~,~ \psi^*_{- \uparrow} \big]^T$$
So that $\check{G}_h(\vec r , {\vec r}~')$ will be arranged into the following matrix,
\bea
\check{G}_h(\vec r, {\vec r}~')=
\begin{pmatrix}
[\hat{G}_{++}]_h & [\hat{G}_{+-}]_h\\
[\hat{G}_{-+}]_h & [\hat{G}_{--}]_h
\end{pmatrix}.
\eea 
In the above equation $[\hat{G}_{\chi\chi'}]_h$ is of the following form,
\bea
\label{General form of G}
\\
\nn
[\hat{G}_{\chi\chi'}]_h=
\begin{pmatrix}
G_{\chi\chi'}^{\downarrow\downarrow}(z,z') & G_{\chi\chi'}^{\downarrow\uparrow}(z,z')\\
G_{\chi\chi'}^{\uparrow\downarrow}(z,z') & G_{\chi\chi'}^{\uparrow\uparrow}(z,z')\\
\end{pmatrix}
e^{[i k_x (x-x')+i k_y(y-y')]},
\eea
where every element in the above equation is obtained from the corresponding element of the
electron Green's function by appropriate replacements of the angles as discussed above. 
After this replacement (and of course changing the sign of energy) the spin-off-diagonal 
elements of the holes Green's functions become,
\bea
\label{ansatz}
G_{\chi \chi'}^{ \bar\sigma \sigma}(z,z')&=&C_{\chi \chi'}^{ \bar\sigma \sigma}(z') e^{-(q_\chi+i\chi b_z ) z}\\
\nn
&-&\frac{ \chi(k_x^\chi - i \sigma k_y^\chi) }{8\pi^2(q_\chi+i\chi b_z)}e^{-(q_\chi+i\chi b_z) |z-z'|} \delta_{\chi \chi'},
\eea
whereas the spin-diagonal components are,
\bea
G_{\chi \chi'}^{\sigma \sigma}(z,z')&=&  \frac{\varepsilon+ i \chi \sigma \partial_z - \sigma b_z}{\chi ( k_x^\chi - i\sigma k_y^\chi) } G_{\chi \chi'}^{\bar{\sigma} \sigma }(z,z'),
\eea
where $k_{x(y)}^\chi =k_{x(y)}+\chi b_{x(y)}$. 
The value of these matrix elements is the same as those for electrons, except 
for the replacement $\sigma\to -\sigma$. 

Up to this point the above expressions are valid for any BC. 
For $\check M_1$-type BC we have:
\bea
\label{G}
\nn
C_{\chi \chi'}^{\bar{\sigma} \sigma}&=&
\frac{\varepsilon+ i \chi \sigma q_\chi - 2 \sigma b_z - \chi e^{ i \sigma \theta_\chi} (k_x^\chi - i \sigma k_y^\chi)}{\varepsilon- i \chi\sigma q_\chi - \chi e^{ i \sigma \theta_\chi} (k_x^\chi - i\sigma k_y^\chi)}\\
      &\times& \frac{\chi (k_x^\chi - i\sigma k_y^\chi)}{8\pi^2(q_\chi + i \chi b_z)}
      e^{-(q_\chi+i \chi b_z)z' } \delta_{\chi \chi'}.
      \eea
where $\theta_-=\Lambda+\pi$ and $\theta_+=\xi+\pi$, and for $\check M_2$-type BC,
for chirality-off-diagonal and chirality-off-diagonal, respectively, we obtain,
\bea
C_{\bar{\chi} \chi}^{\bar{\sigma} \sigma}
=\frac{-i \chi \sigma  (k_x^\chi- i\sigma k_y^\chi) (k_x^{\bar{\chi}}- i\sigma k_y^{\bar{\chi}})}{8\pi^2D_{\chi \chi}^{\bar{\sigma} \sigma}} 
 e^{-(q_\chi+i \chi b_z)z'}, 
\eea
\bea
C_{\chi \chi}^{\bar{\sigma} \sigma}
=\frac{\chi  (k_x^\chi- i\sigma k_y^\chi)}{8\pi^2(q_\chi + i \chi b_z)} 
\big( \frac{N_{\chi \chi}^{\bar{\sigma} \sigma}}{D_{\chi \chi}^{\bar{\sigma} \sigma}} \big) 
 e^{-(q_\chi+i \chi b_z)z'}, 
\eea
where
\bea
D_{\chi \chi}^{\bar{\sigma} \sigma}&=&\bar{\chi} e^{i\chi \theta_{\bar{\sigma}}}(\varepsilon- i \chi \sigma q_\chi ) (k_x^{\bar{\chi}} - i \sigma k_y^{\bar{\chi}})\\
\nn
&-& \chi e^{i\chi \theta_{\sigma}}(\varepsilon- i \bar{\chi} \sigma q_{\bar{\chi}} ) (k_x^{\chi} - i \sigma k_y^{\chi}),
\eea
and
\bea
N_{\chi \chi}^{\bar{\sigma} \sigma}= D_{\chi \chi}^{\bar{\sigma} \sigma}- 2 i  \sigma e^{i\chi \theta_{\bar{\sigma}}}  (q_\chi + i \chi b_z)   (k_x^{\bar{\chi}} - i \sigma k_y^{\bar{\chi}}),
\eea
with $\theta_\uparrow = -\beta$ and $\theta_\downarrow = - \alpha$.

For practical calculations one has to specialize to a specific coordinate system.
The coordinate system can be chosen in such a way that the Fermi arc lies
along the $k_x$ axis. This does not
harm the generality of approach, as always by appropriate rotation along $k_z$ axis, 
a new coordinate system can be chosen in such a way that the new $k_x$ is along the Fermi arc. 
For details, please refer to Appendix~\ref{kaxis.sec}. 

\section{Proximity with superconductor}

Now, we bring a conventional s-wave superconductor (SC) near the WSM. The bulk Hamiltonian of the SC is:
\bea
H_s=[|\vec k_s|^2/(2m) \hat\kappa_3 + \Delta_s \hat\kappa_1]\otimes \hat\sigma_0,
\eea
where $\vec k_s$ denotes the momentum in the SC, $m$ is the electron mass, $\Delta_s$ is the superconducting gap and  
$\hat\kappa_{(i=0\ldots 3)}$ are the Pauli matrices acting in the particle-hole space.
The coupling between WSM and SC is incorporated by:
\bea
{\cal T}=
\begin{pmatrix}
0   &  \Breve t^{\dagger}    \\
\Breve t  &   0
\end{pmatrix},
\eea
where, considering that the tunneling amplitude $t$ is the same for right handed and left handed electrons, 
and the $4\times 8$ matrix $\Breve t$ is constructed as
$\Breve t=t/2(\check t_+~~\check t_-)$ from $4\times 4$ matrices 
$\check t_\alpha=(\hat\tau_z+\alpha \hat\tau_0 + \hat\tau_1 + i\alpha\hat\tau_y)\otimes \hat\sigma_0$, with $\alpha=\pm$.

Based on Dyson equation, the Green's function of the WSM becomes:
\bea
 {\cal G}_W= {\cal G}^0_W+ \sum_{k_s}{\cal G}^0_W .\Breve t^{\dagger}.\check g_s.\Breve t.  {\cal G}_W,
\label{dyson.eqn}
\eea
where we use the symbols $\check g_s$ to denote $4\times 4$ matrices, ${\cal G}$ for $8\times 8$ matrices.
The superscript $0$ in ${\cal G}^0$ denotes the Green's function in Nambu-space when the tunneling is set to zero. 

Assuming that the superconductivity at the surface of the SC is of the same form as its bulk, and that $\Breve t$ and $\Breve t^{\dagger}$
in Eq.~\eqref{dyson.eqn} are independent of $\vec k_s$, we can perform the sum over $\vec k_s$ to obtain the self energy as~\cite{Rao1}:
\bea
\\
\nn
 \sum_{k_s} \Breve t^{\dagger}.\check g_s.\Breve t=\frac{s}{\sqrt{\Delta^2-\epsilon^2}}(\epsilon \hat\kappa_0 - \Delta \hat\kappa_1) \otimes (\hat\tau_0+\hat\tau_x)\otimes \hat\sigma_0 .
\eea
where $s=\pi \rho_0 t$ with $\rho_0$ the density of states of the superconductor at its Fermi level before
becoming superconductor. 
Substituting this result is Eq.~\eqref{dyson.eqn}, we can drive the Green's function for the surface of the 
WSM in presence of a SC.

\subsection{$\check{M}_1$-type BC}
\begin{figure}[t]
\centering
\parbox{5cm}{
\includegraphics[width=6cm]{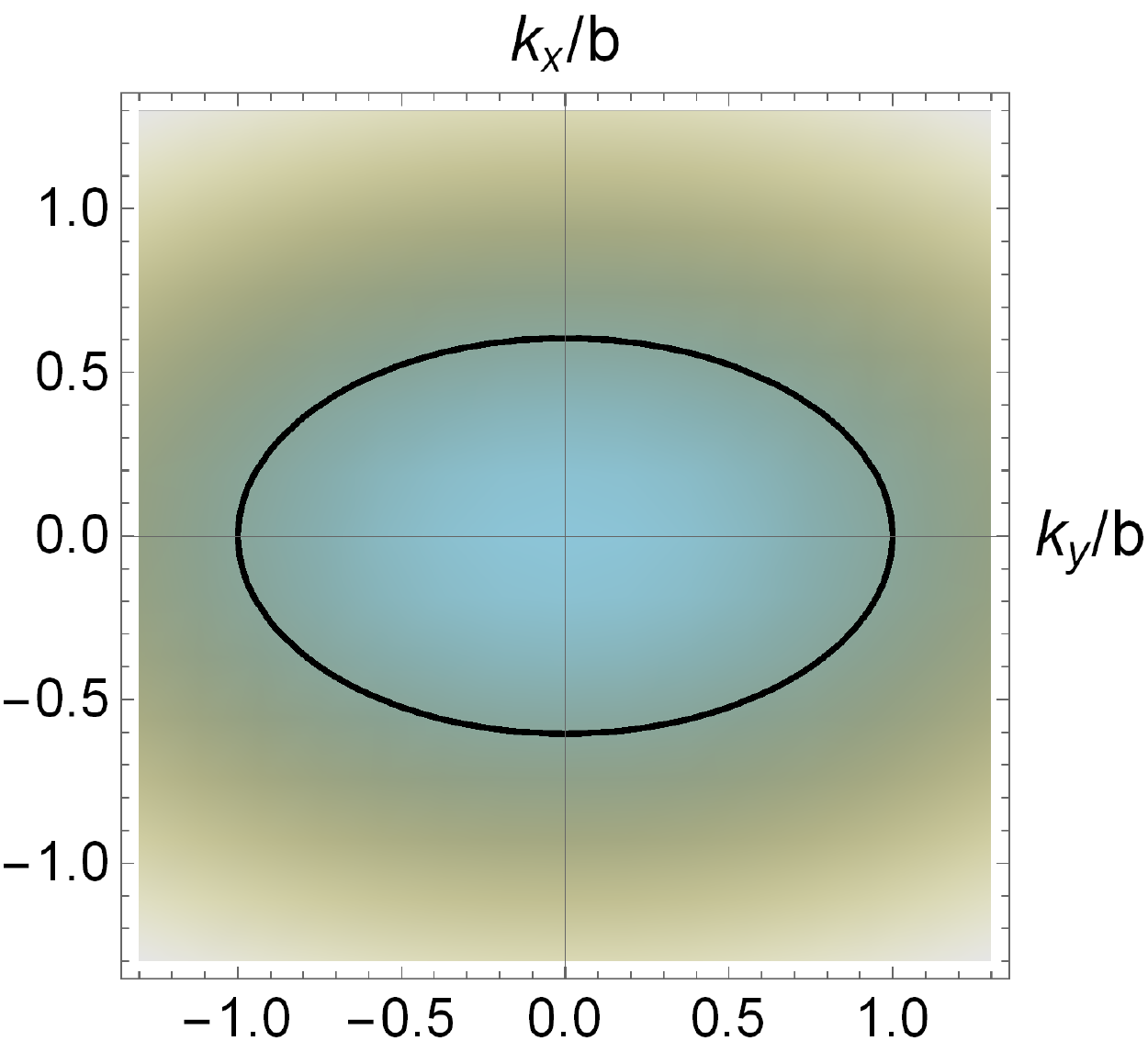}
}
\qquad
\begin{minipage}{2.5cm}
\includegraphics[width=1.1cm]{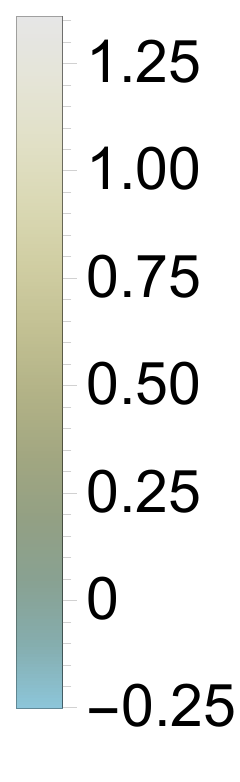}
\end{minipage}
\bigskip 
\caption{Bogoliubov Fermi contour for the first type boundary conditions. The major axis of ellipse coincides with the
Fermi arc of the WSM before bringing the SC to contact with it. The Pfaffian (see the text) changes sign across this contour and excitations
around the elliptic contour are linearly dispersing. The minor axis of the ellipse depends on the tunneling strength as in Eq.~\eqref{ellipse.eqn}. By turning 
off the tunneling the minor axis becomes zero, and ellipse will reduce to the Fermi arc.}
\label{BFC.fig}
\end{figure}

The poles of the Green's function give us the dispersion relation of the excitations on the surface. For $\check M_1$-type BC, 
we obtain the following secular equation for the poles of the Green's function:
\bea
[{\cal F}(\epsilon,\vk) + 4  \epsilon b k_y s^2]^2  -16  s^4  \epsilon^4  (k_x^2+k_y^2)=0
\label{F}
\eea
where ${\cal F}(\epsilon,\vk)=  \sqrt{\Delta^2-\epsilon^2} [4 s^4 (-b^2+k_x^2+k_y^2)- (\epsilon^2-k_y^2)]$ and the tunneling strength $s$
quantifies the ability of electrons in the superconductor to tunnel into WSM. 
The states at the Fermi level correspond to $\epsilon=0$ which will be equivalent to ${\cal F}^2(0,\vk) =0$. Therefore the solutions of ${\cal F}(0,\vk)=0$ will
be twofold degenerate. These solutions are given by the following ellipse in the $k_x-k_y$ plane (see Fig.~\ref{BFC.fig}):
\bea
k_x^2 + (\frac{1+ 4s^4}{4s^4}) k_y^2=1
\label{ellipse.eqn}
\eea 
The major axis of this ellipse is horizontal with magnitude $1$ (note that in our units a momentum of size $1$ actually means $b$) 
and coincides with the Fermi arc of the pristine Weyl semimetal before 
bringing the superconductor to its proximity. 
This is similar to the zero-energy surfaces due to Fermi arcs of {\it doped} WSM~\cite{Haldane_nodal FC}. 
Further, the magnitude of the minor axis,  ${\tilde b}=\frac{2s^2}{\sqrt{1+ 4s^4}}$ is determined by the combination $s$ of the tunneling amplitude $t$
and the density of states $\rho_0$ of the superconductor in its normal phase. As such, when the superconducting agent is
an undoped Dirac superconductor~\cite{Faraei2017}, due to $\rho_0=0$, the minor axis will be of zero length, and the
ellipse will collaps into the Fermi arc. 
It is curious that although the very existence of the ellipse depends on the superconducting gap $\Delta$ of the s-wave superconductor
that proximitizes WSM, the minor axis does not depend on the superconducting gap, $\Delta$ and is only controlled by the tunneling strength $s$.

At $\epsilon=0$, the denominator of Green's function $(\epsilon-H)^{-1}$ will become the determinant of the Hamiltonian, i.e.  
${{\cal F}^2(0,\vec k)}=\det {H(\vec k)}$, where $H(\vec k)$ is the Hamiltoninan of the entire system.
The above relations means that ${\cal F}(0,\vk)$ is actually the Pfaffian of the Hamiltonian. Following Ref. ~\onlinecite{Agterberg1,Agterberg2}
we use ${\cal F}(0,\vk)$ to construct the $Z_2$ topological index $\nu$ that protects the zero-energy ellipse of Bogoliubov quasi-particles as 
$(-1)^\nu=\text{sgn} \big[ {\cal F}({\vec k}_-) {\cal F}({\vec k}_+) \big]$ where ${\vec k}_+ ({\vec k}_-)$ refers to momenta inside (outside) of 
the BFC~\cite{Agterberg1, Agterberg2}.  
As can be seen ${\cal F}(0,\vec k)$ changes its sign across the elliptic zero energy contour 
and therefore we are dealing with $\nu=-1$ situation which is $Z_2$-non-trivial. 
In our two-dimensional case, the $Z_2$ index is only consistent with DIII class which belongs to
BdG family~\cite{Schnyder2008,SchnyderRMP}. In this class, particle-hole and sublattice symmetry
must be present which is the case by construction. The TR must be broken, which is again the case, 
as the parent WSM is characterized by TR breaking parameter $\vec b$. The meaning of $\nu=-1$ is that
weak perturbations within the DIII class are not able to destroy the elliptic Fermi contour of Bogoliubov quasiparticles. 
A simple consequence of this robustness is that by changing the tunneling parameter $s$, only the minor 
axis of the ellipse change, but it can not be cut into pieces or destroyed. As we will see in next sub-section,
with $\check{M}_2$-type BC, we will have a totally different situation. 

In terms of the Altland-Zirnbauer~\cite{Altland-Zirnbauer} classification, the induced superconductivity on Fermi arc
states belongs to the DIII class. The interpretation of its $Z_2$ index is connected with the existence
of (elliptic) BFC. Once the Fermi contour is formed, the Fermi contour {\it itself} as a singularity of 
the Green's function in momentum space
can be further classified by a winding number~\cite{Volovik}. This is defined by 
\be
   n_1=\mbox{tr}\frac{1}{2\pi i}\oint_C G\partial_\ell G^{-1} d\ell
\ee
where the closed path $C$ is any contour enclosing the Fermi contour (ellipse in our case) and $\ell$ parameterizes this path.
For the Fermi contour of two-dimensional metals, as long as it has the Fermi liquid
structure $G(i\omega,p)\propto (i\omega-p)^{-1}$, where $p$ is the momentum deviation from the Fermi contour,
the above winding number will be $\pm 1$. However, an essential difference between the elliptic
Fermi contour of Bogoliubov quasiparticles compared to Fermi contour of Fermi liquids is that, due to two-fold degeneracy, the pole structure near the Fermi contour
is given by $G(i\omega,p)\propto (i\omega-p)^{-2}$. This form of Fermi contour will give $n_1=\pm 2$. This means that in
principle there can be perturbations outside DIII class which can break the $n_1=2$ topological charge into two $n_1=1$ (Fermi liquid-like)
Fermi contours. 

To gain further insight into the physical nature of this BFC, let us study the excitations 
around this elliptic Fermi contour. 
In radial direction, a little away from the ellipse we can use a small parameter $\eta$ to parameterize the momenta at
$\epsilon=0$ as $k_x=(1+\eta) \cos{\phi}$ and $k_y=({\tilde b}+ \eta) \sin{\phi}$. 
Let us assume that by approaching the ellipse, energy vanishes as $\alpha \eta^\gamma$. 
With this choice, the lowest order terms of Eq.~\eqref{F} are:
\bea
\label{leading term}
&& \frac{4 {\tilde b}^4}{1- {\tilde b}^2} \alpha^2 \eta^{2\gamma} {\sin}^2 \phi \\
\nn
&+& \bigg(  \frac{2 \Delta {\tilde b}^2}{1- {\tilde b}^2} \bigg)^2 \eta^2 (\cos^2 \phi +\frac{1}{{\tilde b}} \sin^2 \phi)^2\\
\nn
&+& \bigg[  \frac{8 \Delta {\tilde b}^4}{(1- {\tilde b}^2)^{3/2}} \bigg] \alpha \eta^{\gamma+1} (\cos^2 \phi +\frac{1}{{\tilde b}} \sin^2 \phi) \sin \phi
\\
\nn
&=&  \frac{4  {\tilde b}^2}{1- {\tilde b}^2}  \alpha^4 \eta^{4\gamma} (\cos^2 \phi +\frac{1}{{\tilde b}} \sin^2 \phi)
\eea
If $\gamma>1$, then only the second term on the left-hand side, is the leading order term and should be zero but it is generically impossible. On the other hand if $\gamma<1$, 
then the first term in Eq.~\eqref{leading term} is the leading order term and this leads to $\alpha=0$. We thus conclude that $\gamma=1$ and that around the BFC, 
the energy disperse linearly. There are only two exception to $\gamma=1$: at $\phi=0$ (and $\phi=\pi$ related to the former by symmetry) 
which correspond to dispersion along $k_x$ axis. These two peculiar points correspond to the projection Weyl nodes on the $k_x-k_y$ surface. 
In this case, $\sin\phi=0$ and Eq.~\eqref{leading term} reduces to,
\bea
\bigg(  \frac{ \Delta {\tilde b}^2}{1- {\tilde b}^2} \bigg) \eta^2 =   \alpha^4 \eta^{4\gamma}.
\eea
from which obtain $\gamma=1/2$. Therefore the singular behavior at $\phi=0$ means that
by departing from the projection of Weyl nodes on the $k_x$ direction inward the ellipse, 
we obtain a peculiar $\varepsilon (p_x,p_y=0)\sim \sqrt p_x$ where $p_x$ and $p_y$ measure the momenta
from the two ends of the major axis of the ellipse. 

\subsection{$\check{M}_2$-type BC}
Unlike the $\check{M}_1$-type BC where a robust BFC is obtained which 
can be distorted but not destroyed by changing the parameters of the Hamiltonian (in our case 
the combination $s=\pi\rho_0t$), for $\check{M}_2$-type BC, instead of BFC we will have a set of BW nodes. 
To see this, let us look into the zeros of the determinant appearing in denominator of the Green's function, which
at $\epsilon=0$ becomes,
\bea
\nn
&\bigg\{& \left[ 3(b^2-k_x^2 + ky^2)^2 + 4 k_x^2 (b^2-4k_y+3k_y^2)\right]s^4- b^2 k_y^2 \bigg\}^2
\\
&+&
 16b^2 k_x^2 k_y^2 (2k_y -b)^2 s^4=0
 \label{m2det.eqn}
\eea

\begin{figure*}[t]
\centering
    \centering\includegraphics[width=7cm]{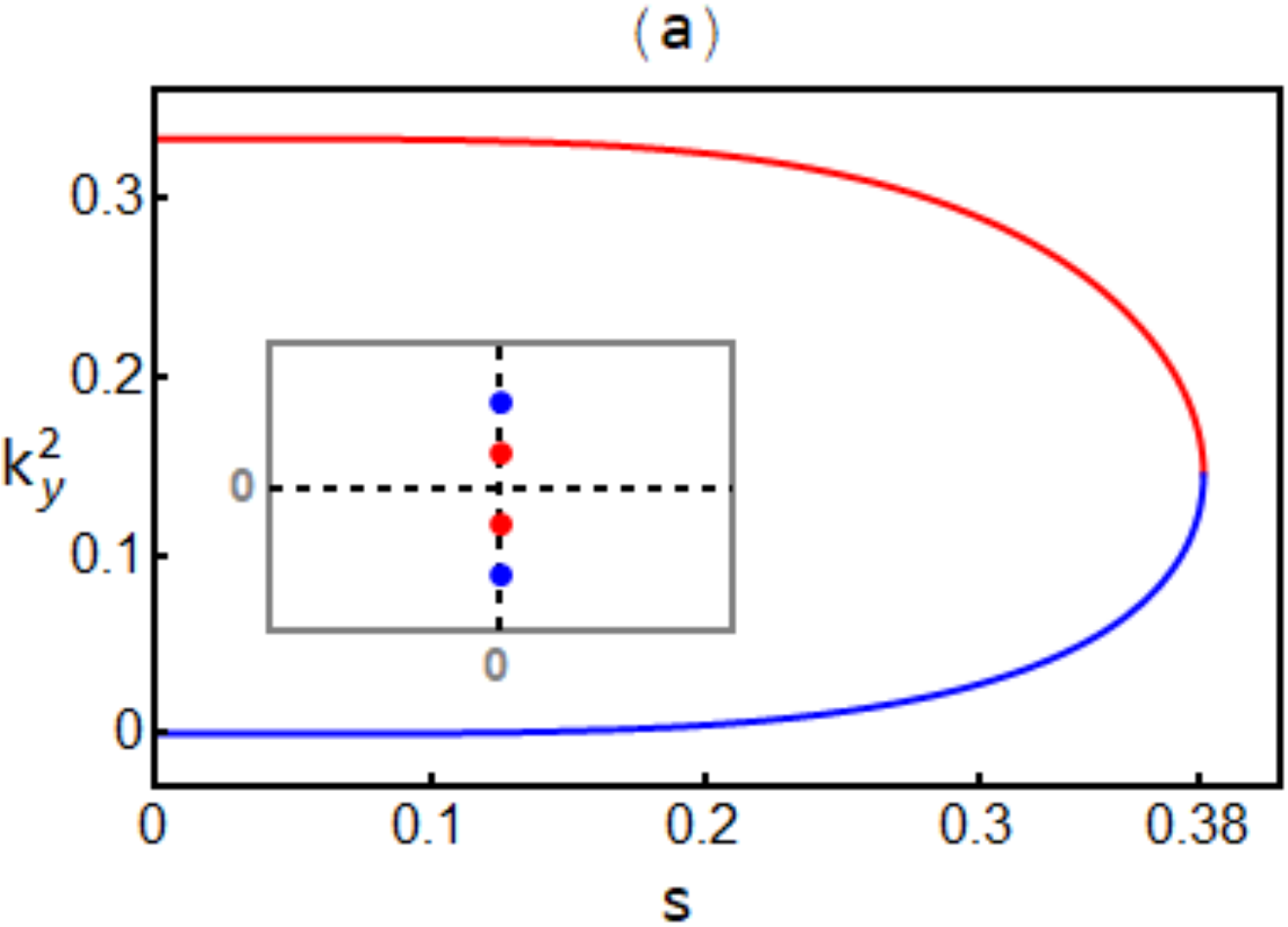}\qquad
    \centering\includegraphics[width=7cm]{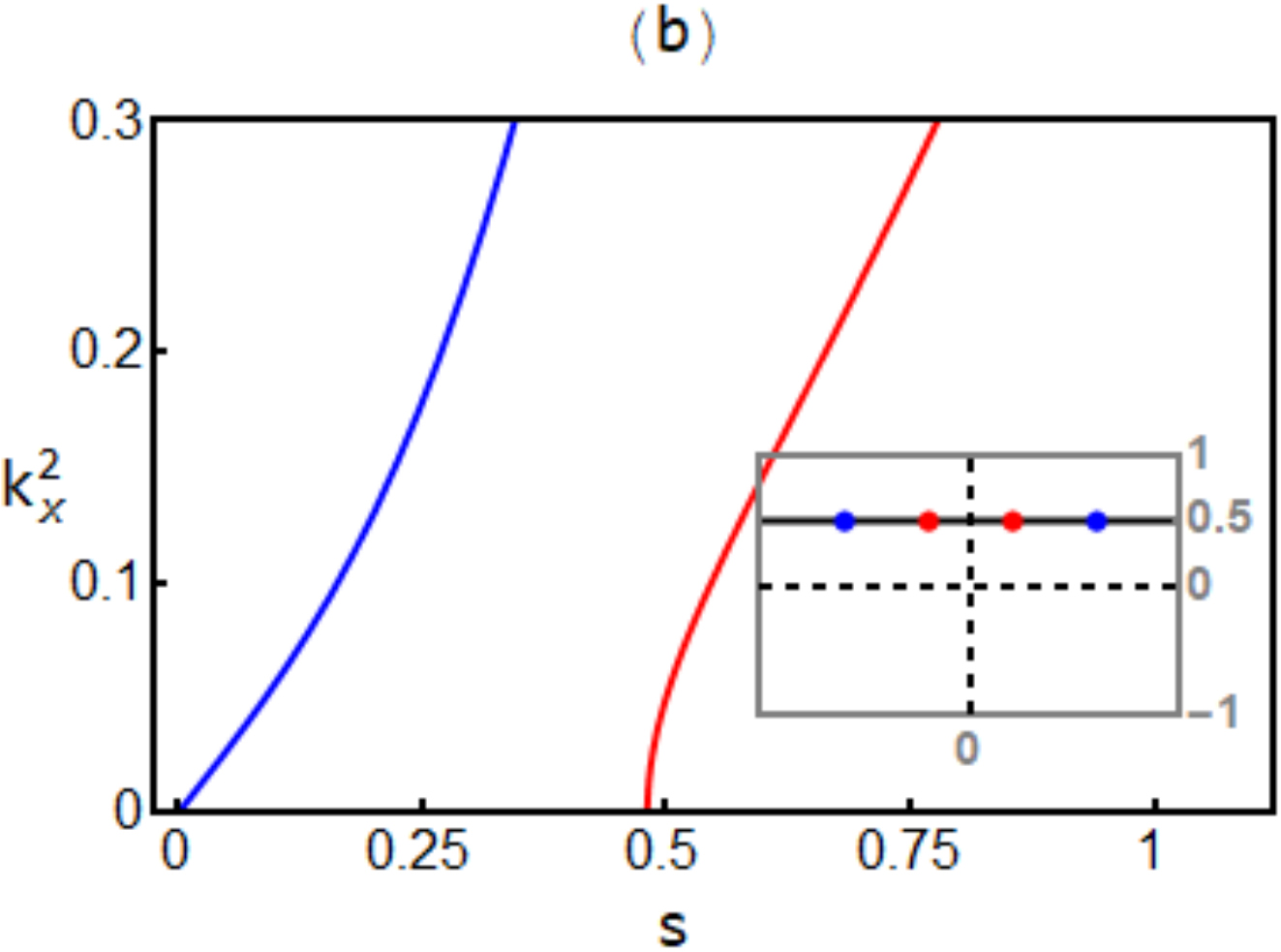}\\
\bigskip
    \centering\includegraphics[width=2.4cm]{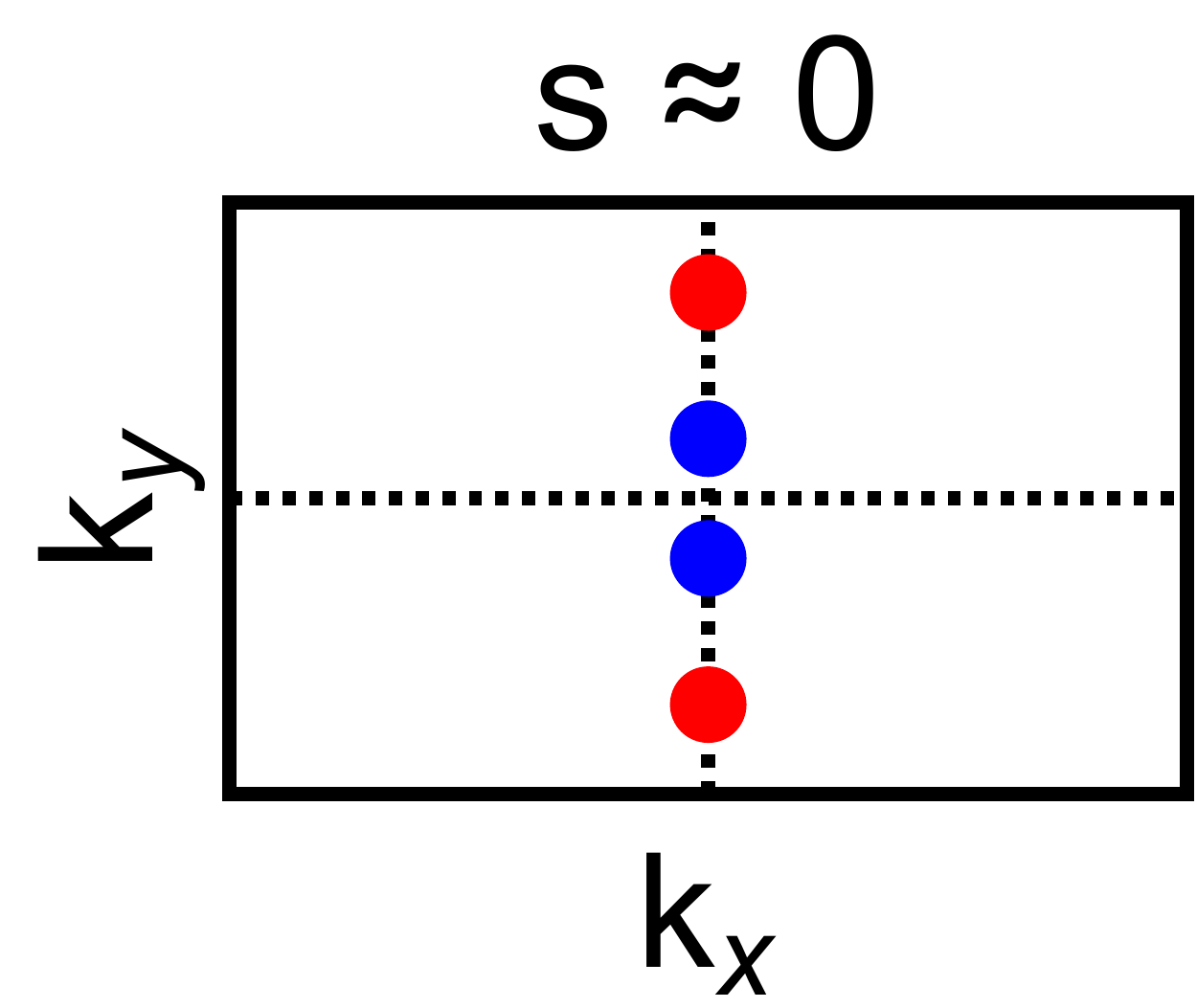}
    \centering\includegraphics[width=2.4cm]{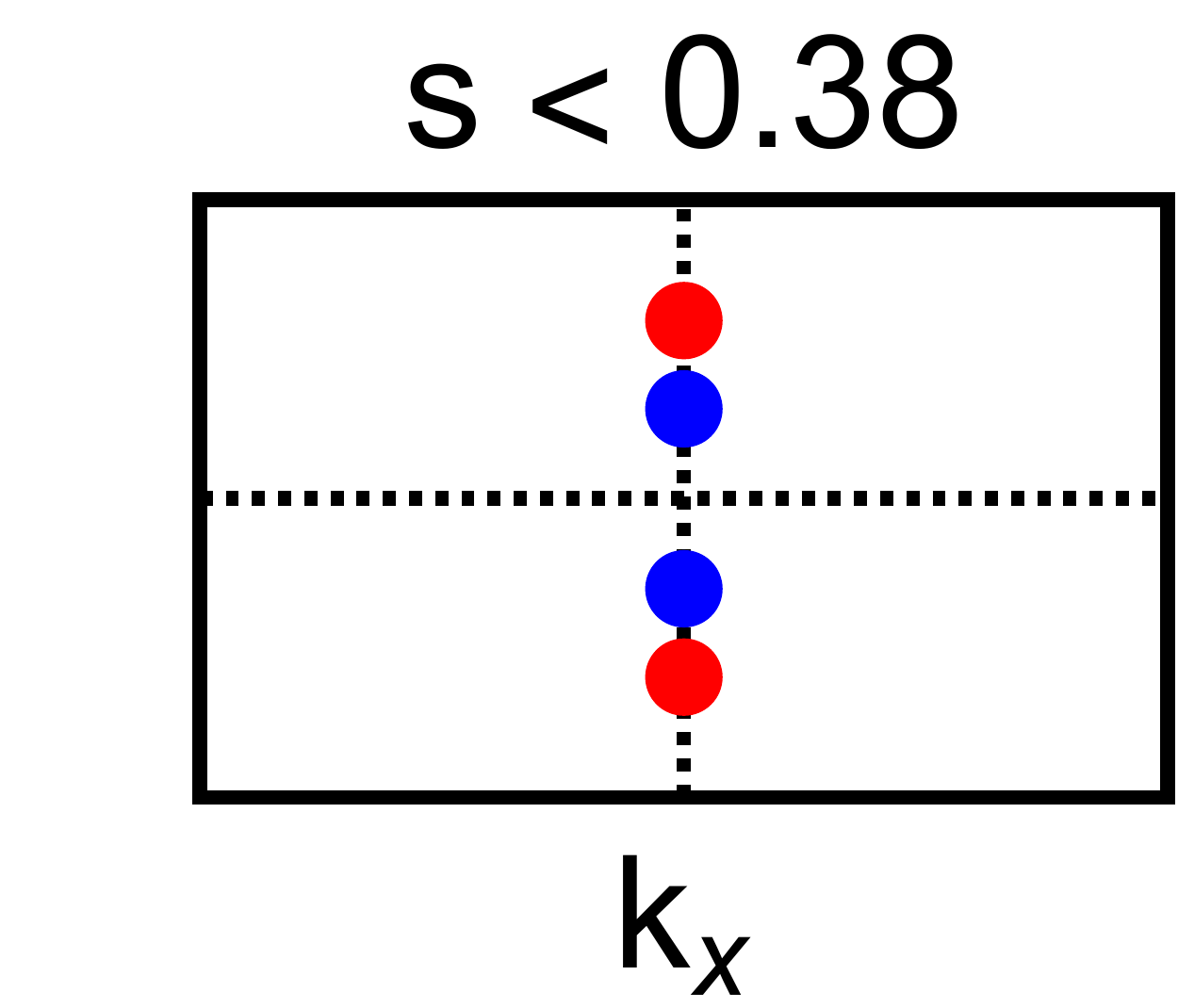}
    \centering\includegraphics[width=2.4cm]{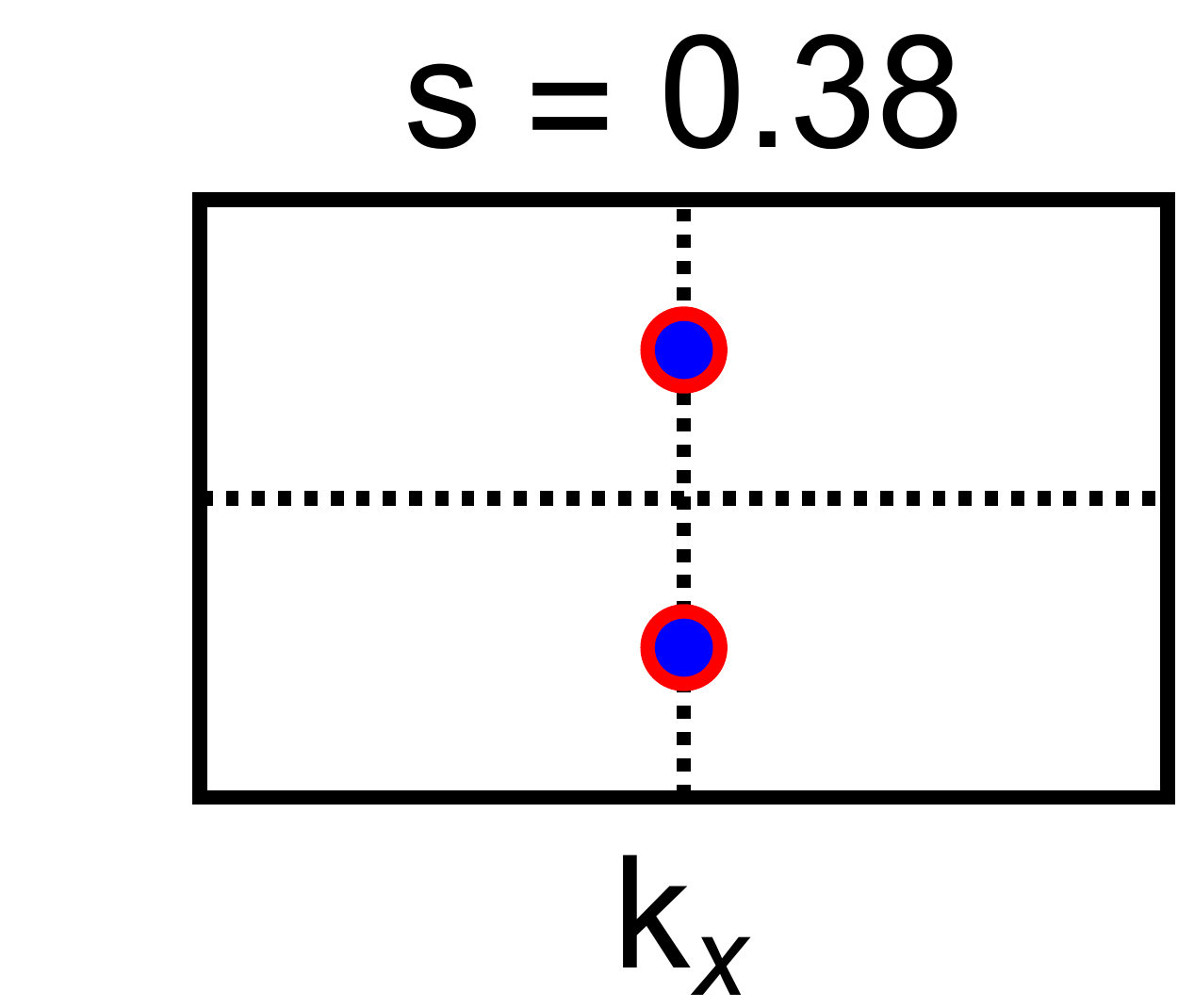}
    \centering\includegraphics[width=2.4cm]{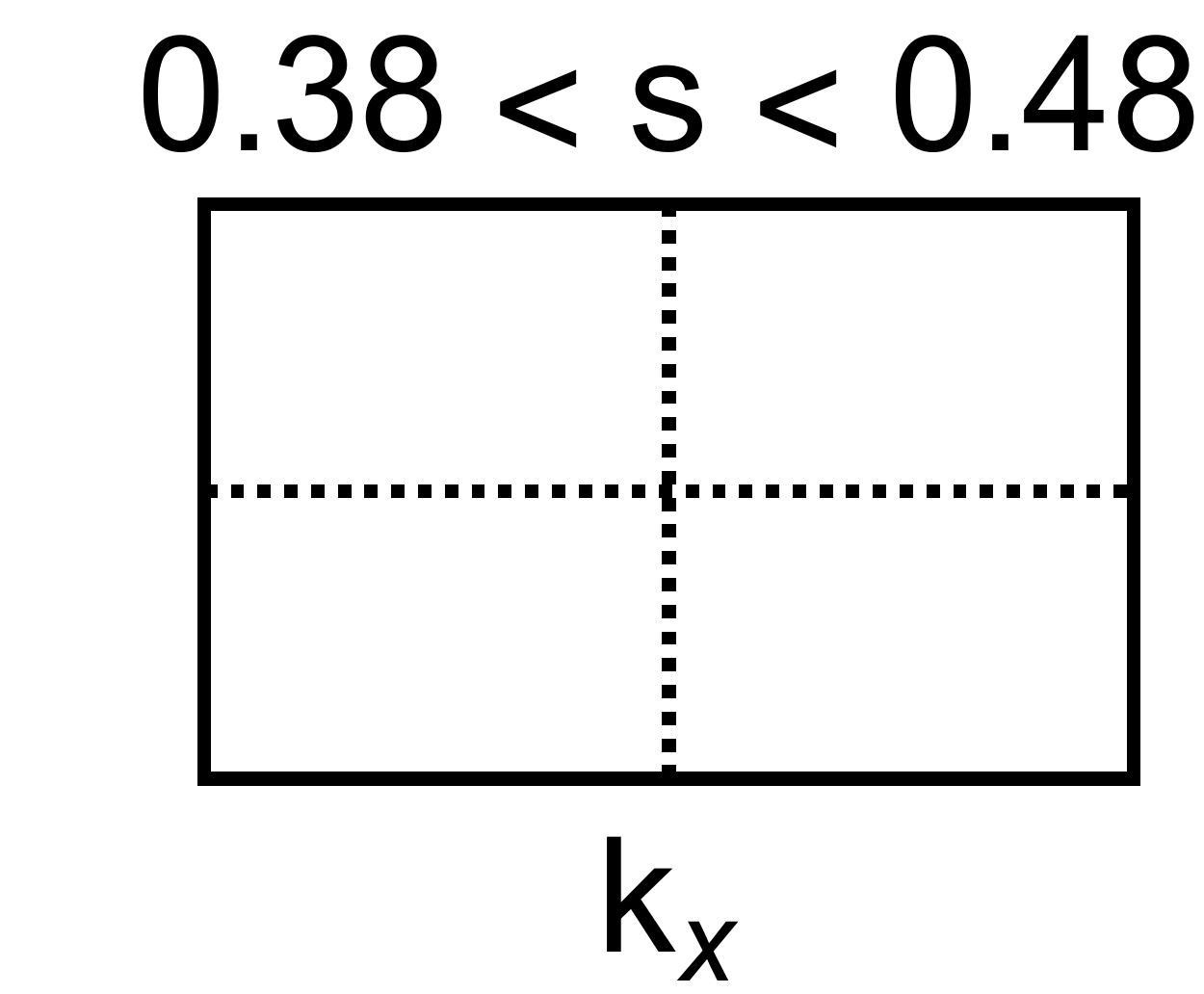}
    \centering\includegraphics[width=2.4cm]{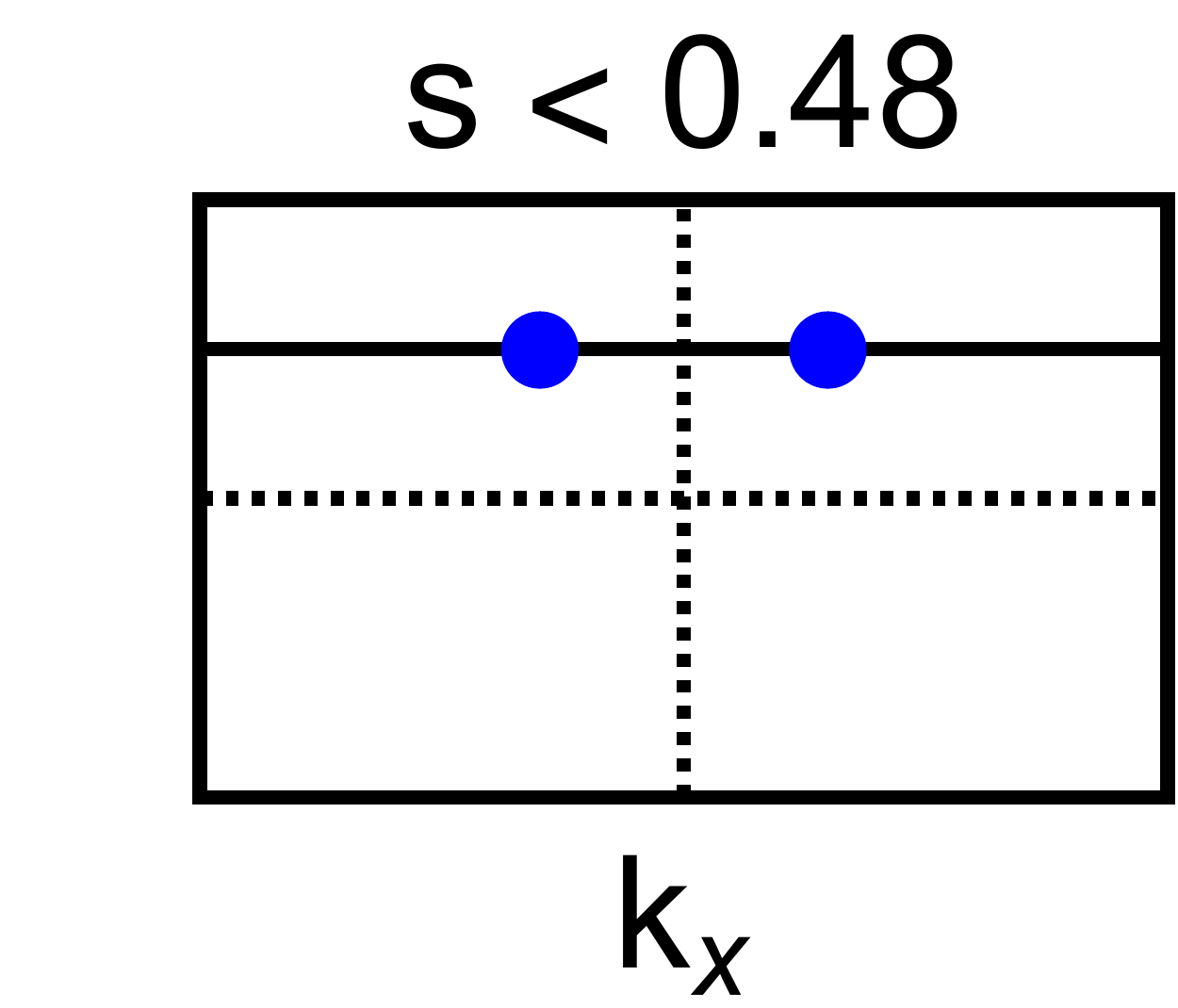}
    \centering\includegraphics[width=2.4cm]{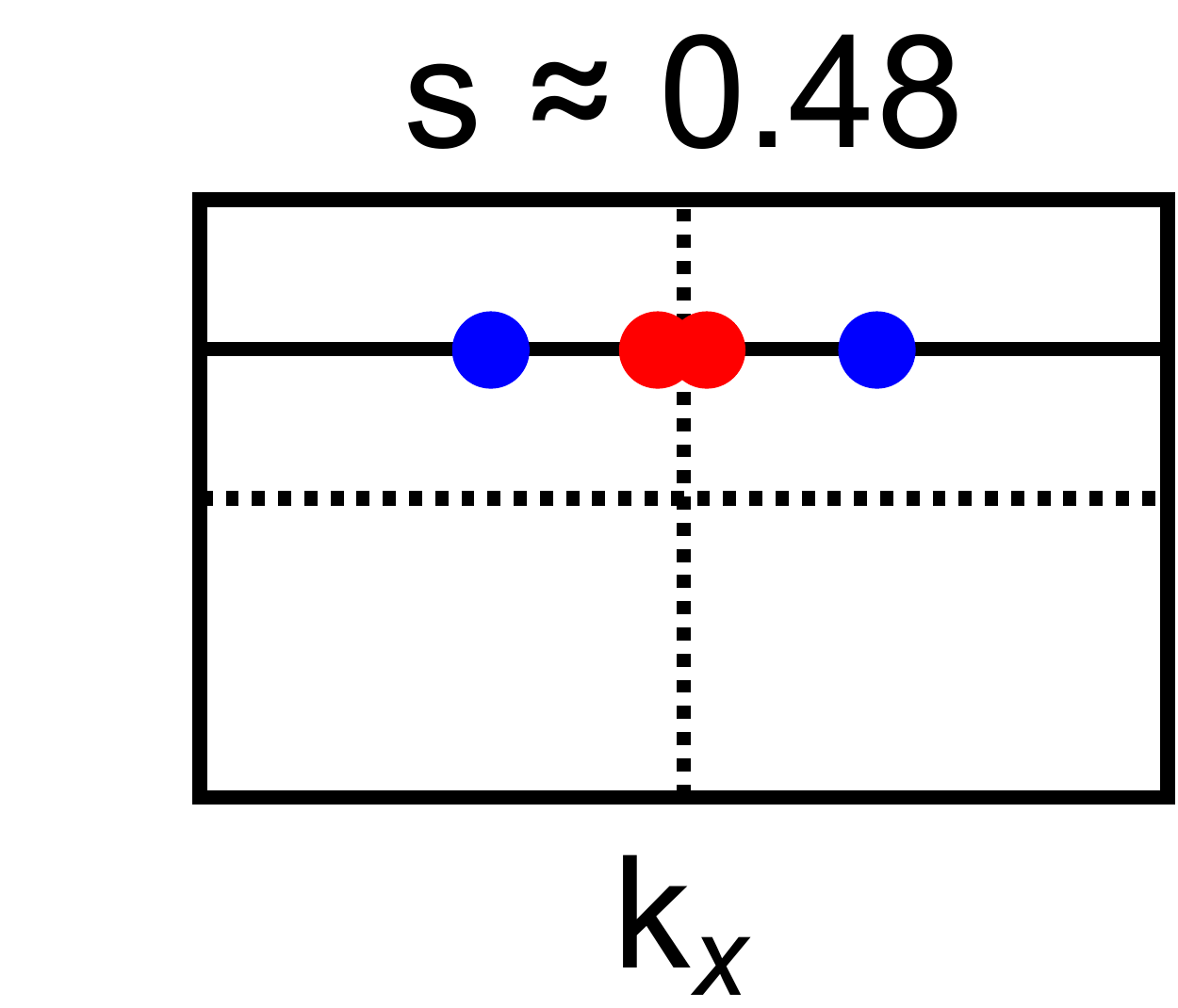}
    \centering\includegraphics[width=2.4cm]{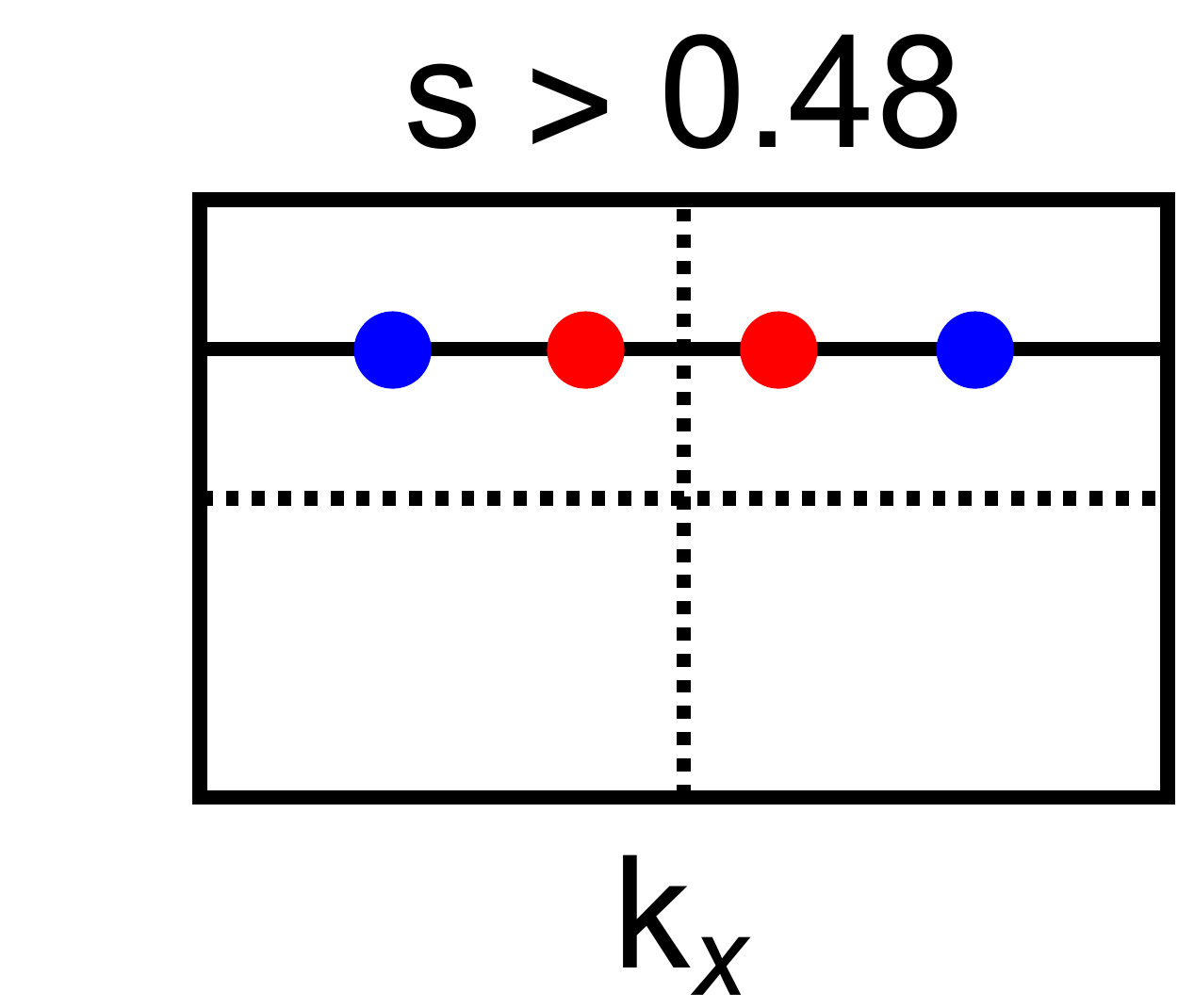}
\caption{Bogoliubov-Weyl nodes with second type boundary conditions. (a) $k_y^2$ coordinate of 
the position of BW nodes as a function of dimensionless tunneling strength $s$. (b) The $k_x^2$ coordinate
of the nodes lying on $k_y=b/2$ as a function of $s$. Inset in both (a) and (b) indicates location of BW nodes. 
Panels in the second row show the schematic evolution of BW nodes upon varying $s$. 
\label{nodes.fig}}
\end{figure*}

This expression being sum of two complete squares appearing in first and second lines, respectively,
can only vanish when each term separately vanishes. From second line there are three possibilities, 
namely $k_x=0$,  $k_y=b/2$, or $k_y=0$. The third case does not give any zero for the first line.
The first two cases, however, give two pairs of solutions as follows (note that we are working in units of $b=1$): 
On the $k_x=0$ line there are two values of $k_y^2$ as long as tunneling is less than $s_{\rm max}=\left[(4-\sqrt{15})/6\right]^{1/4}\approx 0.38$. 
As can be seen in Fig.~\ref{nodes.fig}-a, the two solutions move towards each other and hit at $s_{\rm max}$. Beyond
$s_{\rm max}$ there is no zero energy solution on the $k_y$ axis, meaning that the two BW nodes annihilate each other upon colliding. This indicates that
they are carrying opposite topological charges. 
Their partner in negative $k_y$ axis also behave similarly. This has been schematically shown in the second row of this Fig.~\ref{nodes.fig}. 
On the $k_y=1/2$ line, as can be see in in Fig.~\ref{nodes.fig}-b, the blue pair of BW nodes start at $k_x^2\approx 0$ for very small $s\approx 0$.
As can be seen $k_x^2$ increases linearly as we increase $s$. Beyond $s_{\rm min}=\left(4/75\right)^{1/4}\approx 0.48$, a second
pair of (red) BW nodes appear on the $k_y=1/2$ and start their journey from $k_x^2=0$ point. By further increasing $s$, 
the blue and red BW nodes further depart from each other. 

\section{Pairing symmetry and Majorana Fermi contour}
So far we have shown that $\check{M}_1$-type BC gives a topologically protected BFC. Now we are going to 
discuss its consequences. The Cooper pairs can be either even or odd with respect to its behavior under the exchange of chirality index. 
In the following we separately discuss these two cases. 

\subsection{Even chirality pairing}
\begin{figure}[t]
\includegraphics[width=7cm]{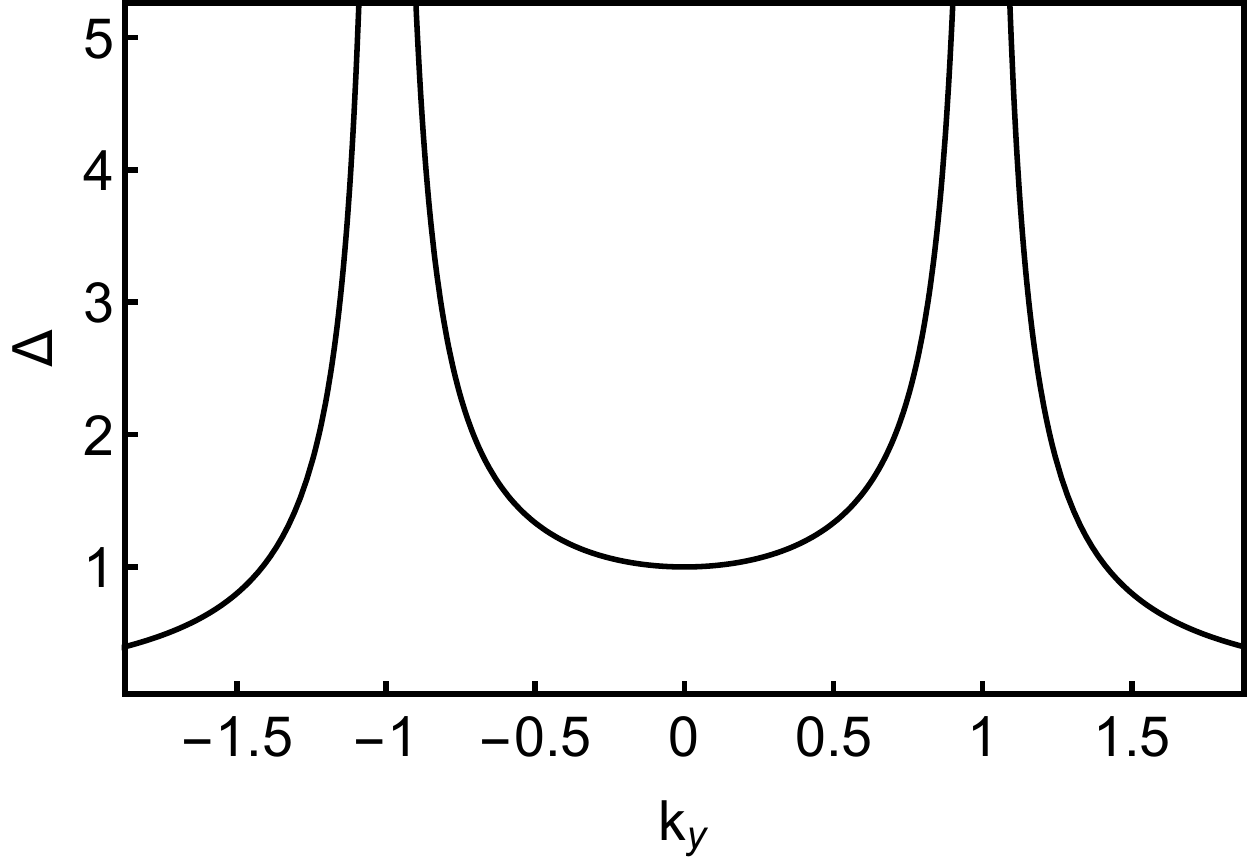}
\caption{The $k_y$ dependence of Eq.~\eqref{integrated.eqn} }
\label{integrated.fig}
\end{figure}

It is useful to form combinations of the pairing amplitudes which are even or odd under exchange of orbital (chirality) index~\cite{Faraei2017}.
Each of these $\Delta$s is a $2\times2$ matrix in spin space and can be written as a sum
of singlet and triplet components as,
\bea
\hat\Delta=i \sigma_y ( d_0 + {\vec d}.\vec\sigma).
\eea
The even interorbital part of the anomalous Green's function  which is even under exchange of band index is given by,
\bea
&&\hat F_+=h
\begin{bmatrix}
-i k_x +k_y & -b\\
-b & i k_x+ k_y
\end{bmatrix},\\
&&
h=\frac{4\Delta_{\rm s} k_y s^2 [{\cal F} - 4 b \epsilon s^2 (\epsilon-k_y)]}{({\cal F} + 4 \epsilon b k_y s)^2-16 \epsilon^2 s^4 k^2}
\eea
which gives, $d_0=0$ and $\vec d= (ik_x,-i k_y,-b)h$. 
The spin-singlet pairing is absent, and therefore the spin angular momentum of the Cooper pairs is even with respect to 
exchange of the spin attribute of the electrons forming the Cooper pair. 
Since the chirality (band index) is already assumed to be even, the orbital part will be necessarily odd. 
It is evident from the $\vec d$ vector that in this channel a substantial $p+ip$ pairing exists. 
However it has been multiplied by a factor $h$ which needs to be integrated over $\epsilon$ to give the
induced pairing. In the weak tunneling regime where $s$
is small, ignoring $s^4$ in comparison to $s^2$ (which also kills the $k_x$-dependence) allows us to analytically
calculate this function which gives the following strength for the pairing, 
\be
   \Delta=-\frac{\pi \Delta_{\rm s} s^2}{\sqrt{|k_y^2-\Delta_{\rm s}^2|}}
   \label{integrated.eqn}
\ee
This function has been plotted in Fig.~\ref{integrated.fig}. On the Fermi arc $k_y=0$, the induced pairing is simply $-\pi \Delta_{\rm s}s^2$ which
conforms to Golden rule intuition. Even on the BFC, according to Eq.~\eqref{ellipse.eqn}, the minor axis is 
controlled by $s^2$, and hence even on the BFC, $k_y$ remains small. 
As can be seen in Fig.~\ref{integrated.fig}, the $k_y$ dependence near $k_y\approx 0$ is very weak,
and therefore this factor will not introduce higher angular momenta and the orbital (angular momentum) part will entirely given by $p+ip$ form. 

The elliptic BFC in our problem is distinct from the underlying Fermi arc. Outside the BFC the Bogoliubov 
quasi particles are more electron-like, while inside the elliptic BFC the excitations are more hole-like.
Right on the BFC the excitations will be equally electron-like and hole-like, so that the average 
charge of the excitations is zero. Therefore the BFC is actually a Majorana Fermi contour. 
The fact that it is protected by a $Z_2$ topological index already manifests as the
simple fact that changing the tunneling strength $s$, does not destroy the elliptic BFC. It
can only modify the aspect ratio and maintains the elliptic shape of the BFC. Now the question will be, what is the
experimental signature of such a Majorana Fermi contour? In a transport setting the portion of
the current which passes through the BFC surface states will appear as a zero-bias
feature. At zero temperature, the strength of such a zero-bias peak is proportional to the perimeter of the Majorana Fermi contour,
\be
   \frac{dI}{dV}\propto 4b E\left(\frac{1}{\sqrt{1+4s^4}}\right)
\ee
where $E$ is elliptic function of second kind, and we have restored the length $2b$ of Fermi arc which
determines the major axis of the ellipse. For low temperatures, the peak will acquire
thermal broadening, but still remains proportional to the above value. According to \onlinecite{Rao_radiation},
the effective length of the Fermi arc can be controlled by coupling to radiation. 
To this extent, the linear dependence of the above formula to the length $2b$ of the Fermi arc can be checked in transport measurements.

The BFC will also have clear thermodynamic signature in the specific heat. Since the two-dimensional BFC
supports linearly dispersing excitations around it (except for two nodal points which are of measure zero),
the resulting density of states will be linear in energy. Therefore the contribution of these excitations
to the specific heat will be $\sim T^2$. This situation is similar to graphene~\cite{Roy}.
This can be pleasantly separated from other degrees of freedom that contribute to absorption of heat. 
First of all, the bulk degrees of freedom of the superconductor have no sub-gap excitations. Secondly the bulk degrees of 
freedom of WSM disperse linearly but in {\it three} space dimensions. By power counting, they will
contribute a $T^3$ term. Therefore the $T^2$ term due to excitations around BFC will take over at low temperatures and can be
separated from the bulk of WSM and SC. 

\subsection{odd-chirality pairing}
The odd amplitude interorbital pairing where is odd under exchange of orbital index is:
\bea
\nn
\hat\Delta_-=\frac{4\Delta k_y s^2 [{\cal f} - 4 b \epsilon s^2 (\epsilon-k_y)]}{({\cal F} + 4 \epsilon b ky s)^2-16 \epsilon^2 s^4 k^2}
\begin{bmatrix}
-i b  & -k_x +i k_y\\
-k_x-i k_y & i b
\end{bmatrix},
\eea
and so $d_0=i k_y h$ and $\vec d= (-i b~,~0~,~k_x)h$. The integration over energy in weak tunneling regime
gives the same formula~\eqref{integrated.eqn}. Although the singlet pairing amplitude $d_0$ is zero on the
Fermi arc ($k_y=0$), nevertheless on the BFC it becomes non-zero value. From Eq.~\eqref{ellipse.eqn}, this value is proportional to the
minor axis $\tilde b\propto s^2$. Therefore the singlet component of pairing on BFC will be controlled by tunneling strength.
On the contrary, the triplet component $\vec d$ of the induced pairing depends on $b$ and $k_x$. The $z$-component of this pairing
changes from $+b$ to $-b$ by spanning the BFC, while its $x$-component remains constant $-ib$. 

\section{Summary and discussion}
We have discussed the proximity induced superconductivity in Fermi arc states.
By chirality blockade, the bulk states play no role in the induced superconductivity in WSM
and the resulting transport is dominated by induced superconductivity in surface Fermi arc states. 
Computing to all-order in tunneling perturbation theory, we find that the original Fermi arc is completely
washed out by coherent all order tunneling of Cooper pairs from the superconductor into WSM. However, 
as a result of this all-order tunneling, a new Bogoliubov Fermi contour is established which is
protected by a $Z_2$ topological index. Such a BFC is actually a Majorana Fermi contour. 
This Majorana Fermi contour shows up as a zero-bias conductance peak, the strength of which is 
proportional to the perimeter of the elliptic BFC. This implies linear dependence on the length $2b$ of the
Fermi arcs. This Fermi contour is protected from
small perturbations. Moreover, in a simple specific heat measurements the BFC at sub-gap temperature scales 
shows up as a distinct $T^2$ contribution to the heat absorption. This can be separated from the $T^3$ contribution from
bulk states of WSM. The bulk of superconductor itself being gapped, is out of game in sub-gap temperature scales. 
By slightly moving away from the Fermi level, the weight of either hole or electron in the Bogoliubov wave
function starts to increase. This might be used for detection of Bogoliubov bands within ARPES or inverse ARPES
measurements. By approaching the Fermi level, the portion of ARPES signal related to projection of 
Bogoliubov states onto hole states will decrease in a characteristic BCS fashion. 

For the second type of BC that flips chirality at the boundary, instead of BFC, we find pairs of
Bogoliubov-Weyl nodes that disperse in the Brillouin zone upon changing the tunneling strength $s$. 
The specific heat signature of Bogoliubov-Weyl nodes is similar to BFC, and goes like $T^2$. The zero-bias
conductance peak for first type-BC is expected to be stronger than those of Bogoliubov-Weyl nodes. 

An interesting question that can be put forward is the following: 
The BFC is a {\it non-interacting} Fermi contour. What happens when strong interactions are included on top of such a Majorana FC
and what are possible gap-opening mechanisms? In the case of $p+ip$ pairing 
a possible strong coupling analogous state can be $\nu=\frac{5}{2}$ quantum Hall state 
which is expected to develop pair density wave gap \cite{CooperClasification}. 

\section{Acknowledgements}
S. A. J. was supported by grant No. G960214 from the research deputy of Sharif University of Technology
and Iran Science Elites Federation (ISEF). Z. F. was supported by a post doctoral fellowship from ISEF.
We are grateful to Mehdi Kargarin for helpful discussions.

\appendix
\section{Matrix elements for Fermi arc along the $k_x$ axis}
\label{kaxis.sec}
Without loss of generality one can rotate the coordinates along $k_z$ axis
in such a way that the Fermi arc will lie along the $k_x$-axis.
This coordinate system corresponds to setting $\Lambda\rightarrow\frac{\pi}{2}$ and $\xi\rightarrow\frac{3\pi}{2}$~\cite{Faraei2017}. 
So, for electrons we have,
\bea
&&k_x^\chi=k_x - \chi b\\
\nn
&&k_y^\chi=k_y\\
\nn
&&q_\chi=-\chi k_x +b\\
\nn
&&\varepsilon=k_y
\eea
and for holes:
\bea
&&k_x^\chi=k_x + \chi b\\
\nn
&&k_y^\chi=k_y\\
\nn
&&q_\chi=\chi k_x +b\\
\nn
&&\varepsilon=-k_y
\eea
After these simplifications, for $\check M_1$-type BC we have,
\bea
&&\big[ G_{\chi\chi}^{\bar{\sigma}\sigma} \big]_e=
\big( \frac{-i\sigma}{4\pi^2} \big)
\frac{k_x-\chi b+i\sigma k_y}{\varepsilon-k_y}
e^{(\chi k_x -b)(z+z')}
\Theta(\bar\chi k_x)
\nn\\
\nn
&&\big[ G_{\chi\chi}^{\sigma\sigma} \big]_e=
\big( \frac{- \chi}{4\pi^2} \big)
\frac{k_x-\chi b+i\sigma k_y}{\varepsilon-k_y}
e^{(\chi k_x -b)(z+z')}
\Theta(\bar\chi k_x)
\\
\nn
&&\big[ G_{\chi\chi}^{\bar{\sigma}\sigma} \big]_h=
\big( \frac{i\sigma}{4\pi^2} \big)
\frac{k_x+\chi b-i\sigma k_y}{\varepsilon+k_y}
e^{(-\chi k_x -b)(z+z')}
\Theta(\chi k_x )
\\
\nn
&&\big[ G_{\chi\chi}^{\sigma\sigma} \big]_h=
\big( \frac{- \chi}{4\pi^2} \big)
\frac{k_x+ \chi b-i\sigma k_y}{\varepsilon+k_y}
e^{(-\chi k_x -b)(z+z')}
\Theta(\chi k_x ).\\
\eea

So the elements of the Green's function matrices for electrons and holes are obtained as follows,
\bea
\\
\nn
&&\big[ G_{++}^{\uparrow\uparrow(\downarrow\downarrow)} \big]_e=
\big( \frac{- 1}{4\pi^2} \big)
\frac{k_x-b\pm i k_y}{\varepsilon-k_y}
e^{( k_x -b)(z+z')}
\Theta_R
\\
\nn
&&\big[ G_{--}^{\uparrow\uparrow(\downarrow\downarrow)} \big]_e=
\big( \frac{1}{4\pi^2} \big)
\frac{k_x+b\pm i k_y}{\varepsilon-k_y}
e^{( -k_x -b)(z+z')}
\Theta_L
\\
\nn
&&\big[ G_{++}^{\uparrow\downarrow(\downarrow\uparrow)} \big]_e=
\big( \frac{\pm i}{4\pi^2} \big)
\frac{k_x- b\mp i k_y}{\varepsilon-k_y}
e^{( k_x -b)(z+z')}
\Theta_R
\\
\nn
&&\big[ G_{--}^{\uparrow\downarrow(\downarrow\uparrow)} \big]_e=
\big( \frac{\pm i}{4\pi^2} \big)
\frac{k_x+ b\mp i k_y}{\varepsilon-k_y}
e^{( -k_x -b)(z+z')}
\Theta_L
\\
\nn
&&\big[ G_{++}^{\uparrow\uparrow(\downarrow\downarrow)} \big]_h=
\big( \frac{- 1}{4\pi^2} \big)
\frac{k_x+  b\mp i k_y}{\varepsilon+k_y}
e^{(- k_x -b)(z+z')}
\Theta_L
\\
\nn
&&\big[ G_{--}^{\uparrow\uparrow(\downarrow\downarrow)} \big]_h=
\big( \frac{1}{4\pi^2} \big)
\frac{k_x- b\mp i k_y}{\varepsilon+k_y}
e^{( k_x -b)(z+z')}
\Theta_R
\\
\nn
&&\big[ G_{++}^{\uparrow\downarrow(\downarrow\uparrow)} \big]_h=
\big( \frac{\mp i}{4\pi^2} \big)
\frac{k_x+ b\pm i k_y}{\varepsilon+k_y}
e^{( -k_x -b)(z+z')}
\Theta_L
\\
\nn
&&\big[ G_{--}^{\uparrow\downarrow(\downarrow\uparrow)} \big]_h=
\big( \frac{\mp i}{4\pi^2} \big)
\frac{k_x- b\pm i k_y}{\varepsilon+k_y}
e^{( k_x -b)(z+z')}
\Theta_R
\eea
and for $\check{M}_2$-type boundary,
\bea
\\
\nn
&&\big[ G_{\chi\chi}^{\bar{\sigma}\sigma} \big]_e=
-\big( \frac{i\chi\sigma}{8\pi^2  b} \big)
\frac{(k_x+i\sigma k_y)^2-b^2}{\varepsilon-k_y}
e^{(\chi k_x -b)(z+z')}
\Theta(\chi k_x < 0)
\\
\nn
&&\big[ G_{\chi\chi}^{\sigma\sigma} \big]_e=
\big( \frac{-1}{8\pi^2  b} \big)
\frac{(k_x+i\sigma k_y)^2-b^2}{\varepsilon-k_y}
e^{(\chi k_x -b)(z+z')}
\Theta(\chi k_x < 0)
\\
\nn
&&\big[ G_{\bar{\chi}\chi}^{\bar{\sigma}\sigma} \big]_e=
\big( \frac{1}{16\pi^2  b} \big)
\frac{(k_x+i\sigma k_y)^2-b^2}{\varepsilon-k_y}
e^{(\chi k_x -b)(z+z')}
\Theta(\chi k_x < 0)
\\
\nn
&&\big[ G_{\bar{\chi}\chi}^{\sigma\sigma} \big]_e=
-\big( \frac{i\chi\sigma}{16\pi^2  b} \big)
\frac{(k_x-\chi b)^2+ k_y^2}{\varepsilon-k_y}
e^{(\chi k_x -b)(z+z')}
\Theta(\chi k_x < 0)
\eea

\bea
\\
\nn
&&\big[ G_{\chi\chi}^{\bar{\sigma}\sigma} \big]_h=
\big( \frac{i\chi \sigma}{8\pi^2  b} \big)
\frac{(k_x- i\sigma k_y)^2-b^2}{\varepsilon+ k_y}
e^{-(\chi k_x +b)(z+z')}
\Theta(\chi k_x >0)
\\
\nn
&&\big[ G_{\chi\chi}^{\sigma\sigma} \big]_h=
\big( \frac{1}{8\pi^2  b} \big)
\frac{(k_x-i\sigma k_y)^2-b^2}{\varepsilon+k_y}
e^{-(\chi k_x +b)(z+z')}
\Theta(\chi k_x >0)
\\
\nn
&&\big[ G_{\bar{\chi}\chi}^{\bar{\sigma}\sigma} \big]_h=
\big( \frac{1}{16\pi^2  b} \big)
\frac{(k_x- i\sigma k_y)^2-b^2}{\varepsilon+k_y}
e^{-(\chi k_x +b)(z+z')}
\Theta(\chi k_x >0)
\\
\nn
&&\big[ G_{\bar{\chi}\chi}^{\sigma\sigma} \big]_h=
\big( \frac{i\chi\sigma}{16\pi^2  b} \big)
\frac{(k_x+\chi b)^2+ k_y^2}{\varepsilon+k_y}
e^{-(\chi k_x +b)(z+z')}
\Theta(\chi k_x >0)
\eea

Expanding the spin components of the matrix we have,
\bea
\\
\nn
&&\big[ G_{++}^{\uparrow\uparrow(\downarrow\downarrow)} \big]_e=
\big( \frac{-1}{8\pi^2  b} \big)
\frac{(k_x\pm i\ k_y)^2-b^2}{\varepsilon-k_y}
e^{( k_x -b)(z+z')}
\Theta_R
\\
\nn
&&\big[ G_{--}^{\uparrow\uparrow(\downarrow\downarrow)} \big]_e=
\big( \frac{-1}{8\pi^2  b} \big)
\frac{(k_x\pm i k_y)^2-b^2}{\varepsilon-k_y}
e^{-(k_x +b)(z+z')}
\Theta_L
\\
\nn
&&\big[ G_{++}^{\uparrow\downarrow(\downarrow\uparrow)} \big]_e=
\big( \frac{\pm i}{8\pi^2  b} \big)
\frac{(k_x\mp i k_y)^2-b^2}{\varepsilon-k_y}
e^{( k_x -b)(z+z')}
\Theta_R
\\
\nn
&&\big[ G_{--}^{\uparrow\downarrow(\downarrow\uparrow)} \big]_e=
\big( \frac{\mp i}{8\pi^2  b} \big)
\frac{(k_x\mp i k_y)^2-b^2}{\varepsilon-k_y}
e^{-(k_x +b)(z+z')}
\Theta_L
\\
\nn
&&\big[ G_{+-}^{\uparrow\uparrow(\downarrow\downarrow)} \big]_e=
\big( \frac{\pm i}{16\pi^2  b} \big)
\frac{(k_x+ b)^2+ k_y^2}{\varepsilon-k_y}
e^{-(k_x +b)(z+z')}
\Theta_L
\\
\nn
&&\big[ G_{-+}^{\uparrow\uparrow(\downarrow\downarrow)} \big]_e=
\big( \frac{\mp i}{16\pi^2  b} \big)
\frac{(k_x- b)^2+ k_y^2}{\varepsilon-k_y}
e^{( k_x -b)(z+z')}
\Theta_R
\\
\nn
&&\big[ G_{+-}^{\uparrow\downarrow(\downarrow\uparrow)} \big]_e=
\big( \frac{1}{16\pi^2  b} \big)
\frac{(k_x\mp i k_y)^2-b^2}{\varepsilon-k_y}
e^{-(k_x +b)(z+z')}
\Theta_L
\\
\nn
&&\big[ G_{-+}^{\uparrow\downarrow(\downarrow\uparrow)} \big]_e=
\big( \frac{1}{16\pi^2  b} \big)
\frac{(k_x\mp i k_y)^2-b^2}{\varepsilon-k_y}
e^{( k_x -b)(z+z')}
\Theta_R
\eea

\bea
\\
\nn
&&\big[ G_{++}^{\uparrow\uparrow(\downarrow\downarrow)} \big]_h=
\big( \frac{1}{8\pi^2  b} \big)
\frac{(k_x\mp i k_y)^2-b^2}{\varepsilon+k_y}
e^{-( k_x +b)(z+z')}
\Theta_L
\\
\nn
&&\big[ G_{--}^{\uparrow\uparrow(\downarrow\downarrow)} \big]_h=
\big( \frac{1}{8\pi^2  b} \big)
\frac{(k_x\mp i k_y)^2-b^2}{\varepsilon+k_y}
e^{( k_x -b)(z+z')}
\Theta_R
\\
\nn
&&\big[ G_{++}^{\uparrow\downarrow(\downarrow\uparrow)} \big]_h=
\big( \frac{\mp i }{8\pi^2  b} \big)
\frac{(k_x\pm i k_y)^2-b^2}{\varepsilon+ k_y}
e^{-( k_x +b)(z+z')}
\Theta_L
\\
\nn
&&\big[ G_{--}^{\uparrow\downarrow(\downarrow\uparrow)} \big]_h=
\big( \frac{\pm i }{8\pi^2  b} \big)
\frac{(k_x\pm i k_y)^2-b^2}{\varepsilon+ k_y}
e^{( k_x -b)(z+z')}
\Theta_R
\\
\nn
&&\big[ G_{+-}^{\uparrow\uparrow(\downarrow\downarrow)} \big]_h=
\big( \frac{\mp i}{16\pi^2  b} \big)
\frac{(k_x- b)^2+ k_y^2}{\varepsilon+k_y}
e^{( k_x -b)(z+z')}
\Theta_R
\\
\nn
&&\big[ G_{-+}^{\uparrow\uparrow(\downarrow\downarrow)} \big]_h=
\big( \frac{\pm i}{16\pi^2  b} \big)
\frac{(k_x+ b)^2+ k_y^2}{\varepsilon+k_y}
e^{-( k_x +b)(z+z')}
\Theta_L
\\
\nn
&&\big[ G_{+-}^{\uparrow\downarrow(\downarrow\uparrow)} \big]_h=
\big( \frac{1}{16\pi^2  b} \big)
\frac{(k_x\pm i k_y)^2-b^2}{\varepsilon+k_y}
e^{( k_x -b)(z+z')}
\Theta_R
\\
\nn
&&\big[ G_{-+}^{\uparrow\downarrow(\downarrow\uparrow)} \big]_h=
\big( \frac{1}{16\pi^2  b} \big)
\frac{(k_x\pm i k_y)^2-b^2}{\varepsilon+k_y}
e^{-( k_x +b)(z+z')}
\Theta_L
\eea

\end{document}